\documentstyle[twoside,fleqn,npb,epsfig,pstricks]{article}
\newcommand{\AmS}{{\protect\the\textfont2
  A\kern-.1667em\lower.5ex\hbox{M}\kern-.125emS}}

\title{Real and virtual photon structure 
       from dijet events}
      
\author{J. Cvach
\address{Institute of Physics, Academy of Sciences 
        of the Czech Republic, \\
        Na Slovance 2, 182 21 Praha, Czech Republic\\
        \vspace*{3mm} Talk given 
        on behalf of the H1 Collaboration
        at the 7th International Workshop on Deep Inelastic Scattering
        and QCD DIS99, Zeuthen, April 19--23, 1999.}%
        \thanks{Supported by GA AV \v{C}R grant no. A1010821.}}

\begin{document}

\begin{abstract}
Jet production in $ep$ collisions is sensitive to the partonic
structure of photon. The latest measurements of dijet production
from the H1 experiment provide new results of the gluon density of real
photons at low $x$ and, for the first time, on the partonic density
of virtual photons. Properties of the photon remnant were measured as
a function of the hard scale defined by the $p_T$ of the jets. 
The comparison of dijet cross sections with the NLO QCD calculation shows
the non-triviality of the concept of virtual photon structure.  
\end{abstract}

\maketitle

\rput[lt](11.5,11.5){\large %
 \begin{tabular}{l}
 {\bf PRA--HEP/99-02} \\ 
 June 1999 \\
 \end{tabular}
 }
\vspace*{-1cm}

\section{INTRODUCTION}

The experiments carried out in the last 15 years at PETRA, PEP, TRISTAN
and recently also at LEP and HERA have shown convincingly 
that due to strong interactions the real photon has a nontrivial
hadron--like structure. In these experiments incoming 
electrons or positrons acted as a source of virtual photons which
were subsequently involved in hard collisions with partons from
another photon or proton. Most of these studies concerned the region of very
small photon virtualities $Q^2$ (typically $\le 1\; \mbox{GeV}^2$ ---
called photoproduction).
With increasing $Q^2$ the situation changes as photon has less time
to develop its hadronic structure. The way the photon structure
varies with $Q^2$ holds important information on the interplay 
between perturbative and non-perturbative aspects of QCD.

Contrary to $e^+e^-$ colliders which are sensitive
via deep inelastic $\gamma^*\gamma$ scattering 
to the quark structure of the real photon,
jet production at the $ep$ collider HERA is sensitive both to gluon
and quark structure of photon.
Moreover the jet cross section at HERA is large enough 
to allow us to investigate the structure not only
of real $Q^2 < 10^{-2}\;\mbox{GeV}^2$ but also of virtual photon
($1.6\le Q^2 \le 80\; \mbox{GeV}^2$ in this paper). All results
reported here were obtained in $e^+p$ collisions
at energies (27.6 + 820) GeV.

\section{GLUON DENSITY OF REAL $\gamma$}

The dijet cross section in $ep$ collisions at HERA is
sensitive to the gluon content of photon especially at lower
$x_\gamma$.\footnote{$x_\gamma$ is the fraction of photon momentum
carried by parton.} To reach $x_\gamma$ as low as 0.04 the 
inelasticity of the reaction
was restricted to $0.5<y<0.7$
in photoproduction data of integrated luminosity $\sim 7.2\;\mbox{pb}^{-1}$.
Jets were obtained by the CDF cone algorithm with $R=0.7$ 
and accepted in the pseudorapidity range
$-0.2<\eta_i<2.5$ in laboratory frame.\footnote{The $+z$ axis of 
the coordinate system points in the direction of incoming proton.} 
The cross section $\mbox{d}\sigma/\mbox{d}x_\gamma$ from
events with at least two jets with $E_{T,i}>6\;\mbox{GeV}$
after pedestal subtraction and $|\eta_1-\eta_2|<1$ was
corrected for detector inefficiencies using PHOJET and PYTHIA.

\begin{figure}[htb]
\vspace{9pt}
\epsfig{file=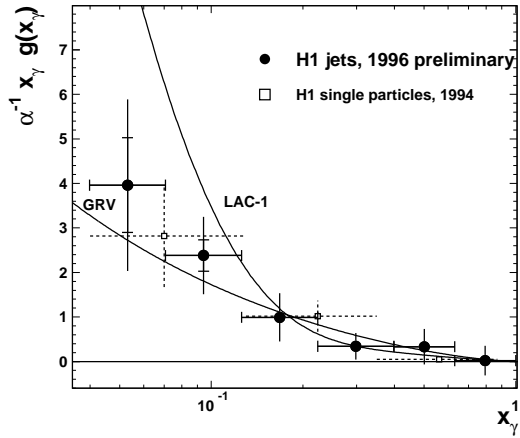,height=65mm,%
bbllx=5pt,bblly=170pt,bburx=550pt,bbury=650pt,clip=}
\caption{Measurement of gluon density in real photon from
dijet ($\bullet$ - this analysis) and
high $p_T$ track ($\Box$) events.}
\label{fig:oliver}
\end{figure}

To obtain the gluon density of the photon the dijet cross section 
was further unfolded to parton level using D'Agostini's program \cite{agostini}. 
The largest systematic error in the resulting effective parton density

\begin{equation}
f_\gamma(x_\gamma,Q^2,E_T^2) = 
\sum_{i}^{N_f}(q_i + {\bar q}_i) + \frac{9}{4} g.
\end{equation}
where $q_i,\bar q_i, g$ are the quark, antiquark and gluon densities in
the photon,
comes from the hadronic $E$ scale uncertainty and different 
treatment of soft parton interactions 
used in PHOJET and PYTHIA. The both dominate at low $x_\gamma$.
The measured effective parton density is almost independent of the hard
scale defined by the $p_T$ of the partons. The gluon density is obtained
from the effective parton density subtracting quark densities as given
by the GRV--LO parametrisation.

The result of \cite{Oliver} is shown in Fig~\ref{fig:oliver} together with 
the H1 analysis of high $p_T$ inclusive tracks \cite{h1gluon}.
Both measurements are consistent with the increase of 
gluon density in real photon at low $x_{\gamma}$. Note 
that the high $p_T$ inclusive track
and the jet analyses are subject to different
detector systematics. The growth of $g(x_\gamma)$ at low $x_\gamma$
agrees with the GRV--LO parametrisation.
  
\section{PARTON DENSITY OF VIRTUAL $\gamma^*$}

The purpose of this analysis is to investigate the evolution
of the effective parton density (1) with $Q^2$ and with 
the hard scale $P_t$. 
Jets were found using the inclusive $k_T$ algorithm.
Events were accepted if the two highest $E_T$ jets in the event
were in the backward hemisphere of the $\gamma^*p$ cms frame with
$|\eta_1-\eta_2|<1, -3<\bar \eta <-0.5$, where $\eta_i$ are jet rapidities and
$\bar \eta$ their average. The constraints on jet $E_T$ were
such that no jet has $E_T<4$ GeV and the sum of jet $E_T$'s
is always $\ge 11$ GeV. The inelasticity of reaction was restricted
to $0.1<y<0.7$ and virtuality of $\gamma^*$ to $1.6<Q^2<80\;\mbox{GeV}^2$.
The integrated luminosity of the data sample is $\sim 6\;\mbox{pb}^{-1}$.
The triple-differential jet cross section 
$\mbox{d}\sigma/\mbox{d}Q^2\mbox{d}\bar E_T^2\mbox{d}x_\gamma$
was measured \cite{tania}, where $ \bar E_T$ is the mean transverse
energy of the two highest $E_T$ jets.
From the cross section a leading order effective parton
density as a function of $x_\gamma \;, Q^2\;, P_t^2$
was extracted.
For unfolding \cite{agostini} the correlation matrix was used
based on HERWIG and RAPGAP Monte Carlo samples.
The largest systematic error $\sim 25\%$ in the parton
density arises from model dependence (hadronization
uncertainty).  
 
\begin{figure}[htb]
\vspace{9pt}
\epsfig{file=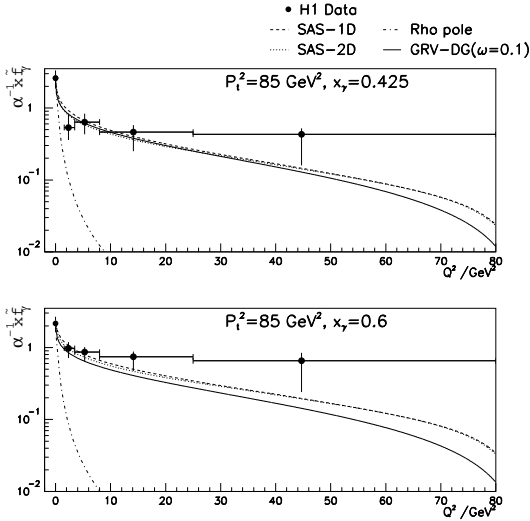,height=75mm}
\caption{The leading order effective parton density of the
photon as a function of $Q^2$ at $P_t^2=85\;\mbox{GeV}^2$. 
The errors are dominated
by systematics. Shown are also GRV and SAS photon parton 
densities and photoproduction data scaled by a $\rho$ pole
factor (dot-dashed curve).}
\label{fig:tania}
\end{figure}

The effective parton density was extracted in the kinematic
range of $0.15<x_\gamma<0.75$ and $30<\bar E_T^2<300\;\mbox{GeV}^2$
in bins of $Q^2, P_t^2$ and $x_\gamma$
with limitation $\langle P_t^2\rangle\! >\!\langle Q^2\rangle$. 
The parton density is approximately independent of $P_t^2$,
and within errors, it is consistent with the normalisation
and logarithmic scaling violations characteristic of the photon
structure.  The logarithmic suppression  with
increasing  $Q^2$ as predicted by virtual photon structure
gives a good description below $Q^2\sim20\;\mbox{GeV}^2$
(Fig~\ref{fig:tania}). Comparison with a $\rho$ pole model shows
that the nonperturbative VDM component plays significant role
only in the parton density of the real photon.

\section{PHOTON REMNANT}

When only a fraction of the photon momentum is involved 
in the hard collision then the remaining momentum is carried away
by spectator partons. These partons fragment into a photon 
remnant which is expected to be approximately collinear with
the original photon. There are at least two possible ways 
to study the properties of the photon remnant. One is to treat it
as an energy deposit close to the direction of the photon,
the other to find it as a low $p_T$ jet by jet algorithms.
In the analysis \cite{remnant}
the $\langle p_t \rangle$ of photon remnant
was measured as a function of the photon virtuality $Q^2<25\;\mbox{GeV}^2$
and $(p_{T,1}+p_{T,2})>12$ GeV of two jets using the DECO \cite{deco} jet 
algorithm. Jets are accepted in pseudorapidity range of
$-3<\eta<0$ in $\gamma^*p$ cms and $x_\gamma<0.75$.

\begin{figure}[htb]
\vspace{9pt}
\epsfig{file=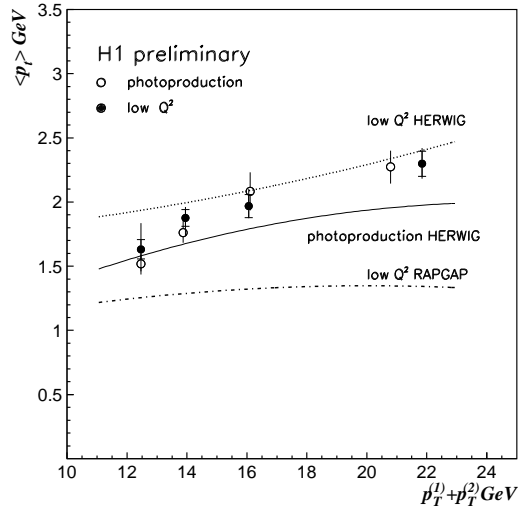,width=75mm,%
bbllx=30pt,bblly=170pt,bburx=520pt,bbury=640pt,clip=}
\caption{Average $\langle p_t \rangle$ of photon remnant for
photoproduction ($\circ$) and low $Q^2 (\bullet$) data as
a function of $p_{T,1}+p_{T,2}$ of two jets.}
\label{fig:alice}
\end{figure}

The $\mbox{d}\sigma/\mbox{d}p_t$ distribution \cite{remnant} 
of the photon remnant
corrected for detector effects agrees better with 
HERWIG than RAPGAP models.
The intrinsic momentum $k_t$ of the parton coming to
the hard process is taken as Gaussian with $k_0=0.66$ GeV to describe
relatively high $\langle p_t \rangle \sim 2$ GeV observed in data. 
It was found that hadronization effects contribute $\sim 1$ GeV
to this value but do not change the $\mbox{d}\sigma/\mbox{d}p_t$ dependence.
Fig~\ref{fig:alice} shows the dependence of average $p_t$ of
the photon remnant on the hard scale given by $p_T$ of jets.
Comparison of photoproduction ($Q^2<0.01$) and low $Q^2$ 
($1.4<Q^2<25\;\mbox{GeV}^2$) data shows 
no significant dependence on the
photon virtuality. On the other hand $\langle p_t \rangle$
of the photon remnant is correlated with the hard scale of
the process ---
it increases with $p_T$ of jets. 
 
\section{COMPARISON WITH NLO}

For the virtual photon with $Q^2\ge 1\;\mbox{GeV}^2$ the concept of
photon structure is in principal not needed since a part of its effects
is included in NLO diagrams. 
Therefore the
question arises whether in NLO the photon structure function is still needed
for the description of data.
To answer this question the dijet cross section was compared
with NLO calculation of JETVIP \cite{jetvip}.

Jets were obtained by the CDF cone algorithm applied to data  of
integral luminosity of $\sim 1\;\mbox{pb}^{-1}$ in
the range $1.4<Q^2<25\;\mbox{GeV}^2$.
Two jets with the highest $E_T$ had to fulfil conditions $E_{T,1}>7,
E_{T,2}>5$ GeV in the pseudorapidity range $-3<\eta_{1,2}<0$ 
in the $\gamma^*$p cms.

\begin{figure}[htb]
\vspace{9pt}
\epsfig{file=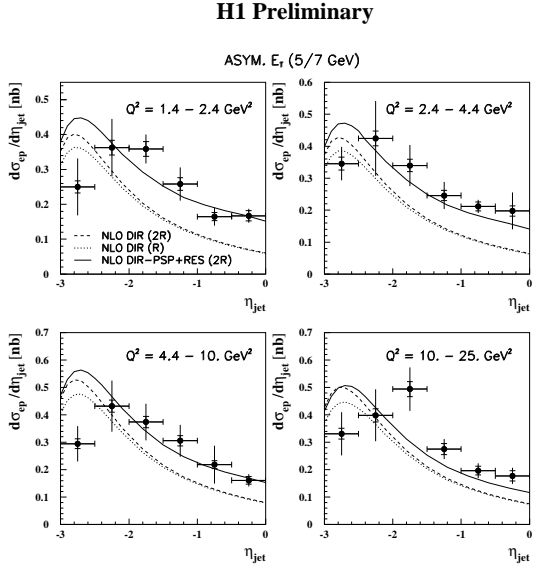,width=70mm,%
bbllx=22pt,bblly=152pt,bburx=523pt,bbury=714pt,clip=}
\caption{The dijet inclusive cross section as a function of jet 
pseudorapidity $\eta$. Data points ($\bullet$) are compared
to NLO calculation of program JETVIP.}
\label{fig:nlo}
\end{figure}

The $\mbox{d}\sigma/\mbox{d}\eta$ distribution of inclusive
dijet production in four $Q^2$ bins is given in Fig~\ref{fig:nlo}
\cite{marek}. The data are corrected for detector inefficiencies.
The curves represent NLO calculation by JETVIP. The dashed line 
represents the contribution from unsubtracted NLO direct processes
(no structure function of the virtual photon --- for details
see \cite{nlo}). The full line
corresponds to the full NLO calculation including also 
diagrams from resolved processes with GRV-HO parton 
distribution functions. We see that for $\eta>-2.5$ and $Q^2<10\;\mbox{GeV}^2$
the unsubtracted direct NLO calculation is systematically below
the data. On the other hand, the full NLO calculation (solid curve)
is in a reasonable agreement with data in all four $Q^2$ bins.

Since hadronization corrections are not included in the NLO
parton level calculation, deviations from the measured cross section
are expected. Estimates show that they increase significantly
at  $\eta<-2.5$ being at the level of $\sim 10\%$ at larger
pseudorapidity.

\vspace*{5mm}     
{\bf \noindent Acknoledgements:}
Thanks to all those who contributed to the analyses presented here.
I want to thank to J. Ch\'{y}la for helpful discussions.
The financial support from the DIS99 Organising Committee
is gratefully acknowledged.

\end{document}